# The luminosity-volume method : Derivation of the cosmological number density in depth from V/V$_m$ distribution
## [Number density in depth from luminosity-volume]

*by*

Dilip G. Banhatti
School of Physics, Madurai Kamaraj University, Madurai 625021, India

**Abstract / Summary.** The classical cosmological V/V$_m$-test is introduced and elaborated. Use of the differential distribution p(V/V$_m$) of the V/V$_m$-variable rather than just the mean <V/V$_m$> leads directly to the cosmological number density without any need for assumptions about the cosmological evolution of the underlying (quasar) population. Calculation of this number density n(z) from p(V/V$_m$) is illustrated using the best sample that was available in 1981, when this method was developed. This sample of 76 quasars is clearly too small for any meaningful results. The method will be later applied to a much larger cosmological sample to infer the cosmological number density n(z) as a function of the depth z.

**Introduction**
A celestial source of isotropic luminosity L at the distance r has the observed flux density S = L / 4.π.r$^2$. Using a telescope of detection limit S$_0$, this source can be observed out to a maximum distance r$_m$ given by S$_0$ = L / 4.π.r$_m^2$. We can associate two volumes with the source: the volume V = 4.π.r$^3$ / 3 actually "occupied" by the source, and the maximum "luminosity-volume" V$_m$ = 4.π.r$_m^3$ / 3 that the source could occupy and still be detected by the telescope at its detection limit S$_0$. The variable x ≡ V/V$_m$ = (r / r$_m$)$^3$ characterizes the fraction of available volume occupied by the source: 0 ≤ x ≤ 1. If the observer is surrounded by a distribution of celestial sources which has uniform density per unit volume relative to r, then x or V/V$_m$ is uniformly distributed on [0, 1]. Conversely, a uniform V/V$_m$-distribution implies a uniform number density (per unit volume) as a function of the distance r from the observer. Testing this for a given sample of N celestial sources may be called the luminosity-volume or V/V$_m$-test, although historically only the mean <V/V$_m$> and the standard deviation σ$_{<V/Vm>}$ of the mean were tested against the population mean <V/V$_m$>$_{pop}$ = ½ and σ$_{<V/Vm>}$ = 1 / √(12.N) (Schmidt 1968, 1978, Schmidt et al 1988, Lynds & Wills 1972, Lynden-Bell 1971, Schmitt 1990).

**The Uniform Random Variable**
In general, for a continuous random variable x, uniform on [0,1],

<x> = ½, σ$_x^2$ = 1/12, {σ$_{<x>}^2$ = 1 / (12.N) for a sample of size N,} and

<(x - <x>)$^{2.r}$> = 1 / {2$^{2.r}$.(2.r + 1)}, <(x - <x>)$^{2.r +1}$> = 0, for r = 1, 2, ….

The mean <x> of a sample of size N is an unbiased estimate of <x>$_{pop}$ within

$\sigma_{<x>} = 1 / \sqrt{(12.N)}$ with probability 68%, since $<x>$ is normally distributed to a very good approximation..

Mid-range as an unbiased estimate of mean
Another unbiased estimate of $<x>$ is $x_{mid} \equiv (1/2).(x_{least} + x_{greatest})$, whose variance is

$\sigma_{x\text{-}mid}^2 = 1 / \{2.(N + 1).(N + 2)\}$ for a sample of size N.

Clearly, $\sigma_{x\text{-}mid}^2 < \sigma_{<x>}^2$ for N > 2 and $\sigma_{x\text{-}mid}^2 << \sigma_{<x>}^2$ for N >> 2.

So maybe $(V/V_m)_{mid}$ rather than $<V/V_m>$ should be used as an estimate of $<V/V_m>_{pop}$, instead of the usual practice of using $<V/V_m>$. However, an advantage of $<V/V_m>$ is that it is distributed normally (with variance $1 / (12.N)$) to a very good approximation for any respectable N (say, N > 10). To consider $(V/V_m)_{mid}$ instead, since its variance is less, we must find its (approximate) distribution, that is, asymptotic distribution as N → ∞ (Banhatti 2009c, in preparation).

Median
The above suggestion of preference for the mid-point of the range as an estimate of central tendency of a (random) variable rather than the mean finds resonance in the trend of plotting median values of quantities rather than their means to bring out correlations between them more clearly (Swarup 1975, Kapahi 1975, Swarup, Subrahmanya & Venkatakrishna 1982, Banhatti 1985, Banhatti & Ananthakrishnan 1989).

**Luminosity-distance and Volume**
For cosmological populations of objects (galaxies, galaxy clusters, radio sources, quasars, γ-ray sources, …) the distance measure r must be replaced by the luminosity-distance $\ell(z)$, and is a function of the redshift z of the object. Similarly, the volume of the sphere passing through the object and centered around the observer is $(4.\pi / 3).v(z)$ rather than $(4.\pi / 3).r^3$. Both $\ell(z)$ and the volume $v(z)$ are specific known functions of z for a given cosmological or world model.

**The luminosity-volume Test**
For example, Kulkarni & Banhatti (1983) and Banhatti (1985) applied the luminosity-volume test to the somewhat unusual cosmology implied by Hoyle-Narlikar (1972) conformal gravity. In general, the (monochromatic) luminosity-distance $\ell_\nu(z)$ depends on z through the spectral shape of the (radio) source since the redshift (by definition) shifts light from higher to lower frequencies ν. For the (radio) sources (quasars) generally used for such cosmological investigations, the spectral shape is roughly parametrized by the negative slope -α of a power-law between $S_\nu$ and ν ($\alpha \equiv$ - dlog $S_\nu$ / dlog ν or $S_\nu$ proportional to $\nu^{-\alpha}$). In the simplest case, a (radio) source of a given (monochromatic) luminosity $L_\nu$ appears to a fixed observer to become monotonically fainter to zero flux density as it is taken farther and farther away to infinity. But some cosmological models may involve refocusing of light in curved, possibly closed and finite, space, so that there may be a minimum flux density at infinity, or even increasing flux density beyond a critical redshift (Kulkarni & Banhatti 1983). A model can, in principle, even have multiple minima, maxima and / or poles in its $\ell_\nu(z)$ relation. In all such cases, the luminosity-volume test can be applied provide the volume "occupied" by a source and the maximum volume "available" to the source are reckoned appropriately (Longair & Scheuer 1970, Lynds & Wills 1972, Avni & Bahcall 1980, Kulkarni & Banhatti 1983, Banhatti 1985).

**Generalities**
The statistics of the $V/V_m$-variable has been considered in some detail by Van Waerbeke et al (1996) in the context of standard cosmologies and cosmological luminosity evolution of the quasar population, using m-z Hubble diagrams and the spread of the $V/V_m$-variable relative to the cosmological parameters. Our purpose here is to relate the distribution of the $V/V_m$-variable to the number density as a function of the redshift z, as briefly indicated by Kulkarni & Banhatti (1983) and Banhatti (1985), putting in any models only when it is necessary. Of course, to calculate the occupied and available volumes V and $V_m$ via the volume function v(z) and luminosity distance $\ell(z)$ or $\ell_\nu(z)$, a cosmological or world model must be assumed. For a general treatment where $\ell_\nu(z)$ is not monotonic (for a given source luminosity $L_\nu$), see Longair & Scheuer (1970), Kulkarni & Banhatti (1983), Banhatti (1985) and Banhatti (2009e, in preparation). Here, we use a model with monotonic $\ell_\nu(z)$ to illustrate the derivation of the number density n(z) from the distribution p(x) of $x = V/V_m$. To be fully general, one must also use a free-form source spectrum rather than parametrizing it, e.g., by a single parameter α, as is usually done. (See Banhatti 2009f (in preparation) for an attempt at such a free-form treatment.) One could take a more complex parametrized form for the spectrum (say, spectral indices $\alpha_{thin}$, $\alpha_{low}$, $\alpha_{high}$ and low- and high-frequency cutoffs $\nu_{low}$ and $\nu_{high}$), since it is the next best approximation after a single spectral index $\alpha \equiv \alpha_{thin}$. (See Banhatti 2009g (in preparation) for such an attempt.) However, our main aim here is to illustrate the relation between the number density n(z) and the (differential) distribution p(x) of $x = V/V_m$.

**Calculation of the Limiting Redshift $z_m$**
For a (radio) source of (monochromatic radio) luminosity $L_\nu$, monochromatic flux density $S_\nu$, (radio) spectral index $\alpha$ ($\equiv$ - dlog $S_\nu$ / dlog $\nu$ so that $S_\nu$ is proportional to $\nu^{-\alpha}$), and redshift z,

$L_\nu = 4.\pi.\ell_\nu^2(\alpha, z).S_\nu$.

For a survey limit $S_0$, the value(s) of $z_m$ is (are) given by

$\ell_\nu^2(\alpha, z) / \ell_\nu^2(\alpha, z_m) = S_0 / S_\nu \equiv s$, $0 \leq s \leq 1$, for a source of redshift z and spectral index α.

This becomes clear on writing the Luminosity $L_\nu$ in terms of $S_0$ and $z_m$ as

$L_\nu = 4.\pi.\ell_\nu^2(\alpha, z_m).S_0$, and comparing or identifying the two expressions for $L_\nu$.

For simplicity, restrict attention to only those cosmological models in which

$[\ell_v(\alpha, z) / \ell_v(\alpha, z_m)]^2 = s$ has a single finite solution $z_m$ for given $\alpha$, $z$ and $S_v$, $S_0$, for the (radio) source under consideration. In other words, we restrict to those models for which $\ell_v(\alpha, z)$ is monotonic increasing with $z$, and $\ell_v(\alpha, 0) = 0$ & $\ell_v(\alpha, \infty) = \infty$. For a sample of flux density limit $S_0$, choosing sources of constant $z_m$ means, for the same $\alpha$, choosing constant $\ell_v(\alpha, z_m) = (S_v / S_0)^{1/2}$. $\ell_v(\alpha, z)$ proportional to $L_v^{1/2}$. Then, for different values of $\alpha$, this amounts to choosing different $L_v(\alpha)$.

Explicitly, for Hoyle-Narlikar (1972) conformal gravity,

$$(H_0 / c)^2 . \ell_v^2(\alpha, z) = z^2 / (1 + z)^{1 - \alpha} \text{ and } (H_0 / c)^3 . v(z) = \{z / (1 + z)\}^3$$
: HN model (Kulkarni & Banhatti 1983).

The function $\ell_v(\alpha, z)$ is the same as the model $q_0 = \sigma_0 = k = 1$, $\lambda_0 = 0$ in von Hoerner's (1974) notation, which may be called the (1, 1, 1, 0) model. <u>The other function $v(z)$, relevant for luminosity-volume test, is **not** the same</u> (cf. von Hoerner 1974).

For the (1, 1, 1, 0) model, which also applies to Hoyle-Narlikar (1972) conformal gravity, as shown by Kulkarni & Banhatti (1983), who complement Canuto & Narlikar (1980) in cosmological tests of HN model,

$$\ell_v(\alpha, z) \begin{cases} \text{is monotonic increasing from 0 to } \infty \text{ for } \alpha > -1 \\ \text{is monotonic increasing from 0 to 1 for } \alpha = -1 \\ \text{has a maximum at } z_0 = -2 / (1 + \alpha) = 2 / (|\alpha| - 1) \\ \text{ & goes to 0 at } z = 0 \text{ and } \infty \text{ for } \alpha < -1 \end{cases}$$

Observationally, $\alpha > -1$ for any radio frequency of interest and also at optical frequencies. Hence, for the cosmological samples of interest, $\ell_v(\alpha, z)$ is monotonic increasing from 0 (at $z = 0$) to $\infty$ (at $z = \infty$), to which case we have limited ourselves for simplicity.

**Relating n(z) to p(V/V$_m$)**
With a large enough flux density-limited deep sample, one may select (radio) sources within a narrow range of $z_m$, and still have sufficient number to determine the number density $n(z)$ from the differential distribution $p(x)$ or $p(V/V_m)$ (Banhatti 2009h, in preparation). Until such very large and deep samples are available, sources of different $z_m$ must be combined together to get a large enough sample to derive $n(z)$ sensibly.

Let $N(z_m).dz_m$ represent the number of (radio) sources of limiting redshifts between $z_m$ and $z_m + dz_m$ in the sample being considered, which covers solid angle $\omega$ of the sky, so that $4.\pi.N(z_m) / \omega$ is the total number of sources of limit $z_m$ per unit $z_m$-interval. Since the total volume available to sources of limit $z_m$ is $V(z_m) = (4.\pi / 3).(c / H_0)^3.v(z_m)$, (where the speed of light $c$ and the Hubble constant $H_0$ together determine the linear scale of the universe,) the number of such sources (per unit $z_m$-interval) per unit volume is

$\{3.N(z_m) / \omega\}.(H_0 / c)^3.(1 / v_m)$, where $v_m \equiv v(z_m)$.

Denote by $n_m(z_m, z)$ the number of sources / unit volume / unit $z_m$-interval at redshift z.

Then, $n(z) \equiv \int_z^\infty dz_m . n_m(z_m, z)$, and

$n_m(z_m, z) = \{3.N(z_m) / \omega\}.(H_0 / c)^3.(1 / v(z_m)).p_m(v(z) / v(z_m))$ for $0 \leq z \leq z_m$,

where $p_m(x)$ is the (differential) distribution of $x \equiv V/V_m$ for a given $z_m$.
For $z > z_m$, $n_m(z_m, z) = 0$, since the sources with limiting redshift $z_m$ cannot have $z > z_m$ (for the type of cosmological model we are considering, viz, with $\ell_v(\alpha, z)$ monotonic increasing from 0 (at z = 0) to ∞ (at z = ∞)). To get the total n(z) for all $z_m$-values, integrate over $z_m$:

$n(z) = \{3 / \omega\}.(H_0 / c)^3. \int_z^\infty dz_m.( N(z_m) / v(z_m)).p_m(v(z) / v(z_m))$.

**The Scheme of Calculation**
The upper limit ∞ for $z_m$ in this integral is illusory, since for a real sample, however deep and large, there will be a maximum $z_{max}$ for the limiting redshift $z_m$. Consequently, $n(z_{max}) = 0$. In fact, the lifetimes of individual (radio) sources will come into the calculation, as well as the galaxy / cluster / structure-formation epoch at some high redshift (say, > 10). Thus, any such n(z) calculation will give useful results only upto a redshift much less than $z_{max}$.
  Formally writing $z_{max}$ instead of ∞ for the upper limit,

$n(z) = \{3 / \omega\}.(H_0 / c)^3. \int_z^{z\_max} dz_m.( N(z_m) / v(z_m)).p_m(v(z) / v(z_m))$ for $0 \leq z \leq z_{max}$.

To apply to real samples, this must be converted into a sum. To this end, divide the $z_m$-range 0 to $z_{max}$ into k equal intervals, each = $z_{max} / k = \Delta z$. The mid-points are

$z_i = (i - ½).\Delta z = \{(i - ½) / k\}.z_{max}$. Calculate n(z) at these points: $n(z_i)$.

Converting the integral into a sum,

$(\omega / 3).(c / H_0)^3.n(z_j) = \sum_{i=j}^k \{N_i / v(z_i)\}.p_i(x_{ij})$, where $x_{ij} = v(z_j) / v(z_i)$.  **(1)**

  It is often more useful / appropriate to use $Z_m = \ln z_m$ as the redshift variable. The integral and the corresponding sum are then:

$n(z) = \{3 / \omega\}.(H_0 / c)^3. \int_z^{z\_max} dZ_m.(z_m.N(z_m) / v(z_m)).p_m(v(z) / v(z_m))$ for $0 \leq z \leq z_{max}$, and

$(\omega / 3).(c / H_0)^3.n(z_j) = \sum_{i=j}^K \{z_i.L_i / v(z_i)\}.p_i(x_{ij})$, where $x_{ij} = v(z_j) / v(z_i)$.  **(1')**

In these two forms (with $z_m$ and $Z_m$ as variables), $N_i$ is the population of the ith $z_m$-bin and $L_i$ that of the ith $Z_m$-bin. There are K bins for the $\ln z_m \equiv Z_m$ variable, and K and k will, in general, be different. Further, the $z_m$- (or $Z_m$-) bins need not all be of the same size. Unequal bins are also allowed / possible and may be more convenient. Since the population M of a bin has uncertainty √M due to counting (or Poisson) statistics, it is advantageous to choose bins so as to have roughly equal numbers of sources each. This way, the error-bars are the same through the range of $z_m$ (or $Z_m$).

**Illustrative Calculation Done in 1981**
Wills & Lynds (1978) have listed a carefully defined sample of 76 optically identified quasars used by Kulkarni & Banhatti (1983) for testing mean $<V/V_m>$ against ½ in a model partially indicated by (1, 1, 1, 0) in von Hoerner's (1974) notation (see above).

Here we use this (small) sample only to illustrate derivation of $n(z)$ from $p(x) \equiv p(V/V_m)$. We use the Einstein-de Sitter cosmology or (½, ½, 0, 0) world model, for which

$$(H_0 / c)^2 . \ell_v^2(\alpha, z) = 4.(1 + z)^\alpha / \{\sqrt{(1 + z)} - 1\}^2 \text{ and } (H_0 / c)^3 . v(z) = 8.\{1 - 1 / \sqrt{(1 + z)}\}^3$$
$$\text{for (½, ½, 0, 0) model (i.e., E-de S)}$$

For each quasar, $z_m$ is calculated by iteration using Newton-Raphson method starting with initial guess z for $z_m$. The values of z, $z_m$ are then used to calculate $v(z)$, $v(z_m)$ and hence $x = V/V_m$. All the 76 $V/V_m$-values are used to plot a histogram. A good approximation for $p(x)$ is $p(x) = 2.x$, which is normalized over [0,1]. The limiting redshifts $z_m$ range from 0 to 3.2. Dividing into four equal intervals, the bins centered at 0.4, 1.2, 2.0 and 2.8 contain 19, 31, 16 and 10 quasars. Although each of these 4 subsets is quite small, we calculated and plotted histograms $p_i(x)$, i = 1, 2, 3, 4 for each subset for x-intervals of width 0.2 from 0 to 1, thus with 5 intervals centered at x = 0.1, 0.3, 0.5, 0.7 and 0.9. Each normalized $p_i(x)$ is also well approximated by $p_i(x) = 2.x$ except $p_4(0.2994)$. So we have done the calculations using this approximation in addition to using the actual values. Finally we calculate $(\omega / 3).(c / H_0)^3.n(z_j)$ using **(1)** (linear scale for $z_m$) and **(1')** (ln, i.e., natural logarithmic, scale for $z_m$, viz., using $Z_m = \ln z_m$ rather than $z_m$). All these calculations are tabulated below.

**Table for $p_i(x)$ and $p(x)$**

| x | No. | $p_1(x)$ | No. | $p_2(x)$ | No. | $p_3(x)$ | No. | $p_4(x)$ | No. | $p(x)$ |
|---|---|---|---|---|---|---|---|---|---|---|
| 0.1 | 0 | 0 | 1 | 0.161 | 0 | 0 | 0 | 0 | 1 | 0.066 |
| 0.3 | 2 | 0.526 | 2 | 0.323 | 3 | 0.9375 | 1 | 0.5 | 8 | 0.526 |
| 0.5 | 3 | 0.789 | 6 | 0.968 | 2 | 0.625 | 1 | 0.5 | 12 | 0.789 |
| 0.7 | 8 | 2.105 | 8 | 1.290 | 7 | 2.1875 | 5 | 2.5 | 28 | 1.842 |
| 0.9 | 6 | 1.580 | 14 | 2.258 | 4 | 1.25 | 3 | 1.5 | 27 | 1.776 |
| Totals | 19 | | 31 | | 16 | | 10 | | 76 | |

**Table of n(z) calculation using linear scale for limiting redshifts**

| j | $z_j$ | $N_j$ | → $v(z_j)$ | i = 1 | i = 2 | i = 3 | i = 4 | → $n(z_j)$ |
|---|---|---|---|---|---|---|---|---|
| 1 | 0.4 | 19 | 2.97E-2 | 1 | 0.1074 | 0.0492 | 0.0321 | 1307. |
| 2 | 1.2 | 31 | 0.27666 | | 1 | 0.4580 | 0.2994 | 255. |
| 3 | 2.0 | 16 | 0.60399 | | | 1 | 0.6536 | 67. |
| 4 | 2.8 | 10 | 0.92407 | | | | 1 | 22. |

In the second table, (a) 5th to 8th columns are $x_{ij}$-values,

(b) → $v(z_j) \equiv (H_0/c)^3 \cdot v(z_j) = 8 \cdot \{1 - 1/\sqrt{(1+z_j)}\}^3$, and
(c) → $n(z_j) \equiv (\omega/3) \cdot (c/H_0)^3 \cdot n(z_j)$.

Use of approximations $p_i(x) = 2.x$ in evaluating the sums **(1)** for each row j = 1, 2, 3, 4 gives virtually the same results. Another table below shows steps in evaluation of n(z) using ln-scale for limiting redshifts, and $p_i(x) = 2.x$, so that no $x_{ij}$-values need be calculated.

**Table of n(z) calculation using ln-scale for limiting redshifts**

| j | $Z_m$-range | mid-$Z_m$ | $z_m$ (i.e. $z_j$) | $L_j$ | → $v(z_j)$ | → $n(z_j)$ |
|---|---|---|---|---|---|---|
| 1 | -1.5to-0.9 | -1.2 | 0.3012 | 7 | 0.015012 | 355. |
| 2 | -0.9to-0.3 | -0.6 | 0.5488 | 11 | 0.060673 | 301. |
| 3 | -0.3to+0.3 | 0.0 | 1.0000 | 27 | 0.201010 | 337. |
| 4 | +0.3to+0.9 | +0.6 | 1.8221 | 23 | 0.530388 | 181. |
| 5 | +0.9to+1.5 | +1.2 | 3.3201 | 8 | 1.117620 | 48. |

The number of sources in bin j is denoted $L_j$ for ln-scale (instead of $N_j$ for linear scale).

**Concluding Remarks**
Due to too few sources in the total sample, and even fewer in the subsamples for different limiting redshift ranges, the results of the calculation are only indicative. No conclusion about the distribution of quasars in redshift is warranted at this stage. For more meaningful results, cosmological samples of size at least a few hundred is needed. Statistical errors for the results should also be calculated. This paper details the method for calculating n(z) from a well-defined sample, and illustrates the method fully, using such a sample of quasars. We propose to apply the method for larger samples of different types of cosmological populations (galaxies, galaxy clusters, radio sources, quasars, γ-ray sources, …). Note that the $V/V_m$ method was first developed for examining the distribution of stars in our Milky Way Galaxy. This application of the test has recently been revived for specific types of stars like white dwarfs.

**Acknowledgments**
The work reported evolved out of discussions with Vasant K Kulkarni in 1981. Computer Centre of Indian Institute of Science was used for the calculations done in 1981. The first draft was written up in 2004-2005 in Muenster, Germany. Radha D Banhatti provided, as always, unstinting material, moral & spiritual support. Westfaelische-Wilhelms University of Muenster is acknowledged for use of facilities. University Grants Commission, New Delhi is acknowledged for financial support.

-x0x-